\documentclass[prl,twocolumn,superscriptaddress,showpacs]{revtex4}

\usepackage{times}
\usepackage{amssymb}
\usepackage[dvips]{graphicx}
\usepackage{subfigure}

\begin{document}

\title{Accuracy of a teleported trapped field state inside a single bimodal
cavity}
\author{Iara P. de Queir\'{o}s}
\affiliation{Instituto de F\'{\i}sica, Universidade Federal de Goi\'{a}s, 74.001-970, Goi%
\^{a}nia (GO), Brazil}
\author{Simone Souza}
\affiliation{Departamento de F\'{\i}sica, Universidade Federal de S\~{a}o Carlos,
13565-905, S\~{a}o Carlos (SP), Brazil}
\author{W. B. Cardoso}
\email[Corresponding author: ]{wesleybcardoso@gmail.com}
\affiliation{Instituto de F\'{\i}sica, Universidade Federal de Goi\'{a}s, 74.001-970, Goi%
\^{a}nia (GO), Brazil}
\author{N.G. de Almeida}
\affiliation{Instituto de F\'{\i}sica, Universidade Federal de Goi\'{a}s, 74.001-970, Goi%
\^{a}nia (GO), Brazil}
\affiliation{N\'{u}cleo de Pesquisas em F\'{\i}sica, Universidade Cat\'{o}lica de Goi\'{a}%
s, 74.605-220, Goi\^{a}nia (GO), Brazil.}

\begin{abstract}
We propose a simplified scheme to teleport a superposition of coherent
states from one mode to another of the same bimodal lossy cavity. Based on
current experimental capabilities, we present a calculation of the \textit{%
fidelity} \ that can be achieved, demonstrating accurate teleportation if
the mean photon number of each mode is at most $1.5$. Our scheme applies as
well for teleportation of coherent states from one mode of a cavity to
another mode of a second cavity, both cavities embedded in a common
reservoir.
\end{abstract}

\maketitle

The teleportation phenomenon \cite{bennett} has received increasing
attention, and a number of protocols have been suggested for its
implementation in various contexts, for example running waves \cite{16,17}
and cavity-QED \cite{14}. Experimentally, teleportation has been
demonstrated for discrete variables \cite{28,29,31,Riebe04,Barrett04}, and
for a single mode of the electromagnetic field with continuous variables 
\cite{32,Takei05}. More recently, teleportation between matter and light was
announced \cite{Sherson06}, where matter and light are respectively the
stationary and flying media.

In the realm of cavity QED, schemes for teleportation of two-particle
entangled atomic states \cite{13}, multiparticle entangled atomic states and
also entangled field states inside high$-Q$ cavities \cite{25,geisa,zheng}
have been proposed. Although these foregoing schemes using high$-Q$\
cavities represent advances by simplifying the procedures required to
teleport states of cavity modes, all experiments implemented till now have
involved only a single high$-Q$ cavity, for reasons related to the complex
experimental challenges such as decoherence and the difficulty of
controlling the interactions. Nonetheless, the cavity is undoubtedly an
important scenario for testing fundamentals of quantum mechanics \cite%
{harochegato} as well as for demonstrating quantum information processing 
\cite{harocheRC}; hence experiments involving teleportation - the
cornerstone of universal quantum computation \cite{chiangnature} - are
expected to be reported soon in the context of high$-Q$ cavity. Aiming at
this goal, our group recently proposed a simplified scheme to teleport a
superposition of zero- and one-photon state \cite{geisa2}, which makes use
of only a single bimodal high$-Q$ cavity, the teleportation occurring from
one mode to another inside the high$-Q$ cavity. Pursuing this idea, here we
propose an oversimplified scheme to teleport a trapped field state with
continuous spectra, the \textquotedblleft Schr\"{o}dinger
cat\textquotedblright -like state (SCS). The experimental setup is shown in
Fig.1. As in ref.\cite{geisa2}, our scheme uses only one bimodal cavity,
supporting mode $1$ and mode $2$, and two two-level atoms, comprehending the
circular states $\left\vert g\right\rangle $ and $\left\vert e\right\rangle $%
, plus Ramsey zones and selective atomic state detectors. The Hamiltonian
including the required dispersive interaction between an atom and the
dissipating cavity field is $H=H_{0}+H_{I}$, where 
\begin{eqnarray}
H_{0} &=&\left. \sum\limits_{i=1}^{2}\hbar \omega _{i}a_{i}^{\dagger
}a_{i}+\sum\limits_{k}\hbar \omega _{k}b_{k}^{\dagger }b_{k}\right. 
\nonumber \\
&+&\left. \frac{\hbar \omega _{0}}{2}\sigma _{z}+\sum\limits_{i=1}^{2}\hbar
a_{i}^{\dagger }a_{i}\chi _{i}\sigma _{ee}\right.  \label{1} \\
H_{I} &=&\sum\limits_{k}\hbar \left( \lambda _{1k}a_{1}^{\dagger
}b_{k}+\lambda _{1k}^{\ast }a_{1}b_{k}^{\dagger }\right)  \nonumber \\
&+&\left. \sum\limits_{k}\hbar \left( \lambda _{2k}a_{2}^{\dagger
}b_{k}+\lambda _{2k}^{\ast }a_{2}b_{k}^{\dagger }\right) \right. \text{.}
\label{2}
\end{eqnarray}%
Here $\sigma _{ee}=\left\vert e\right\rangle \left\langle e\right\vert ,$ $%
a_{i}^{\dagger }$ and $a_{i}$ are, respectively, the creation and
annihilation operators for the i$th$ cavity mode of frequency $\omega _{i}$,
and $b_{k}^{\dagger }$ and $b_{k}$ are the analogous operators for the k$th$
reservoir oscillator mode, whose corresponding frequency and coupling with
the mode $i=1,2$ are $\omega _{k}$ and $\lambda _{ik}$. The atom-field
coupling parameter $\chi _{i}=\frac{g^{2}}{\delta _{i}}$ will always be
adjusted to ensure $g^{2}\tau /\delta _{i}=\pi $, where $g$ is the Rabi
frequency, $\tau $\ is the atom-field interaction time, and $\delta
_{i}=\left( \omega _{i}-\omega _{0}\right) $ is the detuning between the
field frequency $\omega _{i}$\ and the atomic frequency $\omega _{0}$. The
evolution outside the cavity occurs with $\chi _{i}=0$. It is important to
note that the last term in Eq.(\ref{1}) involving $\chi _{i}$\ will be
effective only with one mode at a time. Thus, while the interaction of an
atom with mode $1$ ($2$) of the cavity field is taking place, the relative
phase due to dispersive interaction of this atom with mode $2$ ($1$) of the
cavity field will be negligible. This is true provided that the difference $%
\Delta $ between the two modes be large enough. In addition, to simplify our
estimation of the fidelity of the teleported SCS, we will assume that the
atom-field coupling is turned on (off) suddenly at the instant the atom
enters (leaves) the cavity.

The evolution of coherent states governed by the Heisenberg equations
corresponding to Eqs.(\ref{1}-\ref{2}) are given in detail in Ref. \cite%
{iara}. Here, for brevity, we collect only the main results, assuming a
reservoir at absolute zero temperature, which is an excellent approximation%
\cite{harochegato}. The results of interest for modes $j=1,2$ are 
\begin{eqnarray}
a_{1}(t) &=&\sum\limits_{j=1}^{2}u_{1j}(t)a_{j}(0)+\sum\limits_{k}\vartheta
_{1k}(t)b_{k}(0)\text{,}  \label{7a} \\
a_{2}(t) &=&\sum\limits_{j=1}^{2}u_{j2}(t)a_{j}(0)+\sum\limits_{k}\vartheta
_{2k}(t)b_{k}(0)\text{,}  \label{8a}
\end{eqnarray}%
where 
\begin{widetext}
\begin{eqnarray}
u_{11}(t) &=&\exp \left[- \frac{(A+B)}{2}t\right] \left[
\frac{\left( B-A\right) }{\sqrt{\left( B-A\right) ^{2}+4CD}}\sinh
\left( \sqrt{\frac{\left( B-A\right) ^{2}+4CD}{2}}t\right) +\cosh
\left( \sqrt{\frac{\left(
B-A\right) ^{2}+4CD}{2}}t\right) \right]   \label{9a} \\
u_{12}(t) &=& - \exp \left[- \frac{(A+B)}{2}t\right] \left[
\frac{2C}{\sqrt{\left( B-A\right) ^{2}+4CD}}\sinh \left(
\sqrt{\frac{\left( B-A\right) ^{2}+4CD}{2}}t\right) \right] \text{,}
\label{10a}
\end{eqnarray}
\end{widetext}and 
\begin{eqnarray}
A &=&i\left( \omega _{1} + \chi +\Delta \omega _{1}\right) +\gamma _{11}/2
\label{11a} \\
B &=&i\left( \omega _{2} + \chi +\Delta \omega _{2}\right) +\gamma _{22}/2
\label{12a} \\
C &=&i\Delta \omega _{12}+\gamma _{12}/2  \label{13a} \\
D &=&i\Delta \omega _{21}+\gamma _{21}/2\text{.}  \label{14a}
\end{eqnarray}%
The $\gamma _{jj^{\prime }},$ and $\Delta \omega _{jj^{\prime }}$, $%
j,j^{\prime }=1,2,$ as explained in Ref.\cite{iara}, are the damping rates
and the Lamb-shifts for the two modes, obtained through the Wigner-Weisskopf
approximation \cite{scully} $\sum_{k}\frac{\lambda _{kj}^{\ast }\lambda
_{kj^{\prime }}}{s+i\omega _{k}}\rightarrow i\Delta \omega _{jj^{\prime }}+%
\frac{\gamma _{jj^{\prime }}}{2};~\Delta \omega _{j}\equiv \Delta \omega
_{jj}$,$~\ u_{21}(t)$ and $u_{22}(t)$ can be obtained from $u_{12}(t)$ and $%
u_{11}(t)$, respectively, by simply swapping $A$ and $B$; and $\vartheta
_{jk}(t)$ is an unimportant function when the reservoir is kept at zero
temperature. Eqs.(\ref{7a}-\ref{14a}) can be further simplified by assuming
the following experimental parameters, in the microwave domain. For the
field mode damping times, $\gamma _{11}^{-1}=10^{-3}s$ and $\gamma
_{22}^{-1}=0.9\times 10^{-3}s$, corresponding respectively to modes $1$ and $%
2$, whose frequencies obey the relation $\omega _{2}=\omega _{1}+\Delta $,
where $\Delta /2\pi $ can be adjusted in the range $100$ $kHz$ to $2$ $MHz$ 
\cite{Domokos98}. The two-level atom must be prepared in such way that the
frequency $\omega _{0}$ of the atomic transition $|e\rangle \rightarrow
|g\rangle $, when the atom enters the cavity, be detuned from mode $1$ by $%
\delta =\omega _{1}-\omega _{0}$ and fulfilling the condition $g\overline{n}%
<<\delta +\kappa $, where $\kappa $ is the rate of spontaneous emission and $%
\overline{n}$ is the mean photon number in mode 1. This condition thus
implies, for the detuning with mode $2$, $g\overline{n}<<\Delta +\delta
+\kappa $. Experimentally, the atomic frequency can be Stark shifted using a
time-varying electric field to detune the atomic frequency with each mode 
\cite{Domokos98} by the large amount $\Delta $. As an example, let us
consider an experiment setup prepared obeying $\delta \sim 10^{5}Hz$,\ $%
g\sim 10^{4}Hz$, $\Delta \sim 10^{7}Hz$. Then, the interaction with mode 2
(which we are assuming as possessing higher frequency), will be also
dispersive, and when the atom -- mode 1 interaction produces a $\pi $ pulse,
the coherent state in mode two will evolve according to $\left\vert \beta
\right\rangle \rightarrow $\ $\left\vert e^{i\phi }\beta \right\rangle \sim
\left\vert \beta \right\rangle $, with $\phi =g^{2}t/(\Delta +\delta )\sim
0.03$, which we take into account when calculating the fidelity. Further, we
can assume the cross-damping rates $\gamma _{12}$ and $\gamma _{21}$ taking
as maximum values those of each mode separately, i.e, $\gamma _{12}$, $%
\gamma _{21}\sim 10^{3}s^{-1}$ \cite{nemes}. With these assumptions and
taking into account the dispersive interaction in Eq.(\ref{1}), Eqs.(\ref{9a}%
)-(\ref{10a}) are simplified as following. $u_{12}(t)=u_{21}(t)\cong 0$.
When the atom is out of or enters the cavity in the ground state, $%
u_{11}(t)=\exp\left[ \left( -\frac{\overline{\gamma }}{2}-i\omega
_{1}\right) t\right] $ and $u_{22}(t)=\exp \left[ \left( -\frac{\overline{%
\gamma }}{2}-i\omega _{2}\right) t\right]$. When the atom is in the cavity
in the excited state, $u_{11}(t)=\exp \left\{\left[ -\frac{\overline{\gamma }%
}{2}-i(\omega _{1}+\chi )\right]t\right\} $ and $u_{22}(t)=\exp \left[
\left( -\frac{\overline{\gamma }}{2}-i\omega _{2}\right) t\right] $ when the
atom is interacting with mode 1; or $u_{11}(t)=\left[ \left( -\frac{%
\overline{\gamma }}{2} -i\omega _{1}\right) t\right] $ and $u_{22}(t)=\exp
\left\{\left[ -\frac{\overline{ \gamma }}{2}-i(\omega _{2}+\chi )\right]%
t\right\} $ when the atom is interacting with mode 2. Here $\overline{\gamma 
}=\left( \gamma _{11}+\gamma _{22}\right) /2$, and therefore we have the
important result that the damping rate for each of the two modes is simply
the mean damping rate of the two modes.

\textit{Ideal process. }The ideal SCS to be teleported is prepared by
injecting a coherent state $\left\vert \beta \right\rangle _{2}$ into mode $%
2 $, assuming $\lambda _{ik}=0$ in Hamiltonian (\ref{2}). Then a two-level
atom $1$ is laser-excited and rotated in $R_{1}$ to an arbitrary
superposition $C_{+}\left\vert e\right\rangle _{1}+C_{-}\left\vert
g\right\rangle _{1}$. After that, the atom $1$ crosses the cavity, having
being velocity-selected to interact off-resonantly with mode $2$ such that $%
\chi \tau =\pi $, where $\tau $ is the atom-field interaction time. The atom 
$1$ then crosses $R_{2}$, undergoing a $\pi /2$ pulse, and is detected,
inducing a collapse of the cavity field to the even ($+$) or odd ($-$) SCS, $%
C_{+}\left\vert \beta \right\rangle _{2}\pm C_{-}\left\vert -\beta
\right\rangle _{2}$, where $C_{+}$ and $C_{-}$ are unknown coefficients
obeying $\left\vert C_{+}\right\vert ^{2}+\left\vert C_{-}\right\vert ^{2}=1$%
. The $+$ $(-)$ sign occurs if the atom $1$ is detected in the state $%
\left\vert g\right\rangle _{1}$ $\left( \left\vert e\right\rangle
_{1}\right) $. From now on let us suppose that the even SCS has been
prepared. 
\begin{figure}[t]
\begin{center}
\includegraphics[{height=3.3cm,width=6.5cm}]{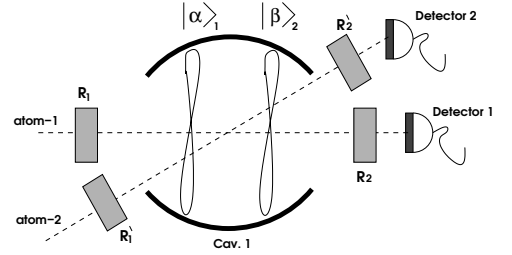}
\end{center}
\caption{Experimental setup for engineering and teleporting a Schr\"{o}%
dinger cat state inside a bimodal cavity. \ The Ramsey zones R$_{1}$, R$_{2}$
(R$_{1}^{\prime }$, R$_{2}^{\prime }$) and atom $1$ (atom $2$) are necessary
for preparing (teleporting) the SCS.}
\end{figure}

The procedure to teleport the SCS is as follows. Firstly, the atom $2$
crosses the Ramsey zone $R_{1}^{\prime }$, undergoing a $\pi /2$ pulse, as
shown in Fig.1, being rotated to the superposition $\sqrt{\frac{1}{2}}\left(
\left\vert e\right\rangle _{2}+\left\vert g\right\rangle _{2}\right) $.
Assuming mode $1$ has previously been prepared in the coherent state $%
\left\vert \alpha \right\rangle _{1}$, the whole state of the system is 
\begin{equation}
\left\vert \varphi \right\rangle =\frac{1}{\sqrt{2}}\left( \left\vert
e\right\rangle _{2}+\left\vert g\right\rangle _{2}\right) \left\vert \alpha
\right\rangle _{1}\left( C_{+}\left\vert \beta \right\rangle
_{2}+C_{-}\left\vert -\beta \right\rangle _{2}\right) \text{.}  \label{3}
\end{equation}%
Next, atom $2$ interacts off-resonantly with mode $1$, such that $\chi \tau
=\pi $, resulting in:%
\begin{eqnarray}
\left\vert \psi \right\rangle &=&\frac{1}{\sqrt{2}}\left[ C_{+}\left\vert
e\right\rangle _{2}\left\vert -\alpha \right\rangle _{1}\left\vert \beta
\right\rangle _{2}+C_{+}\left\vert g\right\rangle _{2}\left\vert \alpha
\right\rangle _{1}\left\vert \beta \right\rangle _{2}\right.  \nonumber \\
&+&\left. C_{-}\left\vert e\right\rangle _{2}\left\vert -\alpha
\right\rangle _{1}\left\vert -\beta \right\rangle _{2}+C_{-}\left\vert
g\right\rangle _{2}\left\vert \alpha \right\rangle _{1}\left\vert -\beta
\right\rangle _{2}\right] \text{.}\phantom{000}  \label{8}
\end{eqnarray}%
Soon after the atom $2$ and mode $1$ interaction, which leads to Eq.(\ref{8}%
), the Stark shift is switched to a large detuned $\delta =\left( \omega
_{a}-\omega _{0}\right) $, thus freezing the evolution corresponding to mode 
$1$ and, at the same time, initiating the atom $2$ and mode $2$ interaction.
The result, after this interaction, is%
\begin{eqnarray}
\left\vert \chi \right\rangle &=&\frac{1}{\sqrt{2}}\left[ C_{+}\left\vert
e\right\rangle _{2}\left\vert -\alpha \right\rangle _{1}\left\vert -\beta
\right\rangle _{2}+C_{+}\left\vert g\right\rangle _{2}\left\vert \alpha
\right\rangle _{1}\left\vert \beta \right\rangle _{2}\right.  \nonumber \\
&+&\left. C_{-}\left\vert e\right\rangle _{2}\left\vert -\alpha
\right\rangle _{1}\left\vert \beta \right\rangle _{2}+C_{-}\left\vert
g\right\rangle _{2}\left\vert \alpha \right\rangle _{1}\left\vert -\beta
\right\rangle _{2}\right] \text{.}\phantom{000}  \label{9}
\end{eqnarray}%
After crossing the bimodal cavity, atom $2$ crosses the Ramsey zone R$%
_{2}^{\prime }$ undergoing a $\pi /2$ pulse, such that Eq.(\ref{9}) evolves
to 
\begin{eqnarray}
\left\vert \vartheta \right\rangle _{2ab} &=&\frac{1}{2}\left[ \left\vert
e\right\rangle _{2}\left\vert -\beta \right\rangle _{2}\left(
C_{+}\left\vert -\alpha \right\rangle _{1}-C_{-}\left\vert \alpha
\right\rangle _{1}\right) \right.  \nonumber \\
&+&\left. \left\vert e\right\rangle _{2}\left\vert \beta \right\rangle
_{2}(-C_{+}\left\vert \alpha \right\rangle _{1}+C_{-}\left\vert -\alpha
\right\rangle _{1})\right.  \nonumber \\
&+&\left. \left\vert g\right\rangle _{2}\left\vert \beta \right\rangle
_{2}(C_{+}\left\vert \alpha \right\rangle _{1}+C_{-}\left\vert -\alpha
\right\rangle _{1})\right.  \nonumber \\
&+&\left. \left\vert g\right\rangle _{2}\left\vert -\beta \right\rangle
_{2}(C_{+}\left\vert -\alpha \right\rangle _{1}+C_{-}\left\vert \alpha
\right\rangle _{1})\right]  \label{14}
\end{eqnarray}

Therefore, by detecting the atom $2$ and measuring the phase of the field in
mode $2$, the field state in mode $1$ is projected on to one of the four
possibilities allowed by Eq.(\ref{14}). Assuming atom $2$ being detected in
its ground state, the phase of the field in mode $2$ can be measured by
injecting a reference field of known amplitude $\beta $ into mode $2$, which
makes the field states $\left\vert \beta \right\rangle _{2}$ and $\left\vert
-\beta \right\rangle _{2}$ in Eq.(\ref{14}) evolve respectively to the
states $\left\vert 2\beta \right\rangle _{2}$ and $\left\vert 0\right\rangle
_{2}$. Such states can then easily be distinguished by sending a stream of
two-level atoms, all of them in the ground state $\left\vert g\right\rangle
_{s}$, to interact resonantly with mode $2$ of the cavity field. Thus, if at
least one of these atoms are detected in their excited state $\left\vert
e\right\rangle _{s}$, indicating the result $\left\vert g\right\rangle
_{2}\left\vert \beta \right\rangle _{2}$ in Eq.(\ref{14}), then mode $1$ is
projected exactly on the desired state $\left\vert \Psi \right\rangle
_{1}=C_{+}\left\vert \alpha \right\rangle _{1}+C_{-}\left\vert -\alpha
\right\rangle _{1}$, thus completing successfully the teleportation process.
On the other hand, if the measurement result is always $\left\vert
g\right\rangle _{s}$, indicating the result $\left\vert g\right\rangle
_{2}\left\vert -\beta \right\rangle _{2}$ in Eq.(\ref{14}), a second atom
interacting off-resonantly with mode $1$ leads to $C_{+}\left\vert -\alpha
\right\rangle _{1}+C_{-}\left\vert \alpha \right\rangle _{1}\rightarrow $\ $%
C_{+}\left\vert \alpha \right\rangle _{1}+C_{-}\left\vert -\alpha
\right\rangle _{1}$. For measurements revealing the states $\left\vert
e\right\rangle _{2}\left\vert -\beta \right\rangle _{2}$ and $\left\vert
e\right\rangle _{2}\left\vert \beta \right\rangle _{2}$ in Eq.(\ref{14}),
the teleportation process cannot be completed unless additional cavities
and/or atoms be introduced, thus overcomplicating the scheme. The
teleportation is accomplished provided we let $\alpha =\beta $, and the
probability of success for the ideal case is then limited to $50\%$.

\textit{Real process. }In real processes, the state $\left\vert \Psi
\right\rangle _{1}$ to be teleported will evolve under the influence of the
reservoir, becoming a mixture $\rho (t)$\ after traced out the reservoir. To
estimate losses in teleportation, we have to compute i) the known value of
the reference field $\beta \left( t\right) $ we have to inject in the cavity
in order to obtain $D\left[ \beta \left( t\right) \right] \left\vert \beta
(t)\right\rangle _{2}=\left\vert 2\beta (t)\right\rangle _{2}$, as remarked
after Eq.(\ref{14}), and ii) the fidelity $\digamma =_{1}\left\langle \Psi
\right\vert \rho (t)\left\vert \Psi \right\rangle _{1}$ \ of the teleported
SCS. To answer question i), we have to compute the evolution $\left\vert
\alpha (0)\right\rangle _{1}\left\vert \beta (0)\right\rangle _{2}\left\vert
\left\{ 0\right\} \right\rangle _{R}\rightarrow \left\vert \Psi
(t)\right\rangle _{12R}$ and then to trace out mode $1$ and the infinite
modes of the reservoir, denoted by $\left\{ 0\right\} $, in order to obtain $%
\left\vert \beta (t)\right\rangle _{2}$. Again we quote the result in \cite%
{iara}: starting from the initial state $\left\vert \alpha (0)\right\rangle
_{1}\left\vert \beta (0)\right\rangle _{2}\left\vert \left\{ 0\right\}
\right\rangle _{R}$ we obtain $\left\vert \beta (0)\right\rangle
_{2}\rightarrow \left\vert \beta (t)\right\rangle _{2}$, where $\beta
(t)\cong u_{22}(t)\beta (0)$, thus answering question i). Note that the
remarkable result that at zero temperature a coherent state loses excitation
coherently remains valid, even when more than one mode is considered. To
answer question ii), we need the evolution of the teleported ideal state $%
\left\vert \Psi \right\rangle _{1}$ in the presence of mode $2$ and the
reservoir,\textit{\ i.e}., we have to calculate the evolution of the
combined state $\left\vert \Psi \right\rangle _{1}\left\vert \beta
(0)\right\rangle _{2}\left\vert \left\{ 0\right\} \right\rangle _{R}=\left(
C_{+}\left\vert \alpha \right\rangle _{1}\left\vert \beta (0)\right\rangle
_{2}\left\vert \left\{ 0\right\} \right\rangle _{R}+C_{-}\left\vert -\alpha
\right\rangle _{1}\left\vert \beta (0)\right\rangle _{2}\left\vert \left\{
0\right\} \right\rangle _{R}\right) $, and, after that, to trace out mode $2$
the and reservoir. This calculation only differs from that in i) by the
second term. The result is the mixed SCS 
\begin{widetext}
\begin{equation}
\rho _{1}(t)=\emph{N}\left\{
\begin{array}{c}
\left\vert u_{11}(t)\alpha _{0}\right\rangle _{11}\left\langle
u_{11}(t)\alpha _{0}\right\vert +\left\vert -u_{11}(t)\alpha
_{0}\right\rangle
_{11}\left\langle -u_{11}(t)\alpha _{0}\right\vert  \\
+\emph{Z(t)}\left[ \left\vert u_{11}(t)\alpha _{0}\ \right\rangle
_{11}\left\langle -u_{11}(t)\alpha _{0}\ \right\vert +h.c.\right]
\end{array}%
\right\}   \label{21}
\end{equation}%
\end{widetext}where $h.c.$ means Hermitian conjugate, $\alpha _{0}=\alpha
(0) $ , and$\ Z(t)=\exp [-2\left\vert \alpha (0)\right\vert
^{2}(1-\left\vert u_{11}(t)\right\vert ^{2})]$ is the term responsible for
decoherence. It is important to note that while $t$ in step i) is the time
spent preparing SCS in mode $2$, in step ii) $t$ is the time after the SCS
is teleported to mode $1$. Restricting ourselves to the joint measurement
corresponding to $\left\vert g\right\rangle _{a}\left\vert \beta
\right\rangle _{2}$, which is the only result promptly leading to
teleportation without requiring additional unitary operations (see Eq.(\ref%
{14})), in about 25\% of the trials the final teleported state will be
exactly the original SCS \textit{provided that we let} $\beta (0)=\alpha (0)$%
. \ According to \cite{Domokos98}, the time the atom spends inside the
cavity is $40-50$ $\mu s$,\ while the total flight time in the experiment is
in the $300-400$ $\mu s$ range, which implies an error around 12\% when, as
usually is done, the time during which the atom interacts with the modes is
neglected. Our calculation, however, is more realistic, since it takes into
account the time the atom crosses the cavity.

In Fig. 2 we present the fidelity for the teleported SCS, calculated with
experimental parameters appropriate for present-day technology \cite%
{harochePRL, harocheRC,Domokos98}. Although these experimental parameters
were quoted from experiments involving orthogonally polarized modes, the
experimental capabilities such as time-varying electric field for
controllable Stark shift, two-level Rydberg atoms, the value of the Rabi
frequency $g$, and so on are expected to work as well for non-orthogonally
polarized modes of a same cavity. For orthogonally polarized modes, apart
from a relative $\pi /2$ phase, each mode will couple with a different
reservoir, and the dynamical fidelity for the prepared and teleported SCS in
a given mode will depend solely on the presence of its corresponding
reservoir, being independent of the second reservoir as well as of the
excitation in the second mode \cite{Iara07}. Also, our scheme applies as
well for teleportation of SCS from one mode of a cavity to another mode of a
second cavity, if both cavities are placed in the same reservoir. In this
last case our scheme will work irrespective of the difference $\Delta
=\omega _{2}-\omega _{1}$. 
\begin{figure}[t]
\begin{center}
\includegraphics{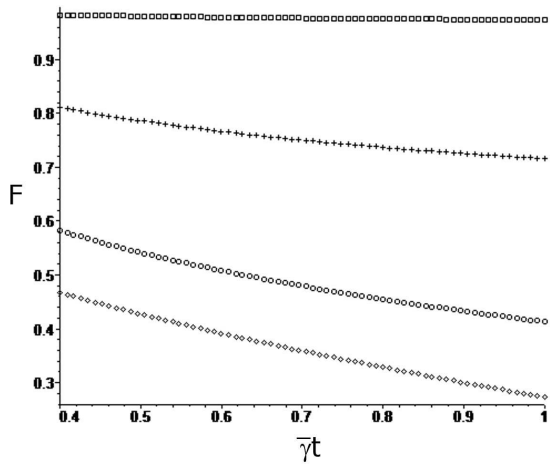}
\end{center}
\caption{Fidelity for the teleported SCS for $\protect\alpha =0.5$, $1.0$, $%
1.5$, and $2.0$. The SCS damping rate for each of the two modes is the mean
damping rate value of the two modes. Here we used the experimental values $%
\protect\gamma _{11}^{-1}=$ $10^{-3}$ $s$ and $\protect\gamma %
_{22}^{-1}=0.9\times 10^{-3}$ $s$ for mode $1$ and mode $2$, respectively.}
\end{figure}

Note from Fig.2 that a successful realization of the teleportation process
is obtained for $\alpha $ ranging from $0.5$ to $1.0$. However, while for $%
\alpha =0.5$ the fidelity remains around unity for all times, for $\alpha
=1.0$\ the fidelity decays to around $0.85$ by the time teleportation is
completed, reaching the lowest value $0.7$ at long times, still a
significantly high value. For $\alpha =1.5$ the fidelity of the SCS by the
time the teleportation is concluded is around $0.6$, higher than the
classical limit $0.5$, showing that the teleported SCS has not been
substantially degraded. On the other hand, our result shows that
teleportation fails for $\alpha \geq 2.0$ given current experimental
capabilities. Note that, although we have calculated a conditional fidelity, 
\textit{i.e.}, the fidelity resulting from 25$\%$ of the trials, it would be
possible to recover, from any of the measurement results, the original SCS
to be teleported, at the expense of introducing additional cavities and/or
atoms. \ However, this procedure would demand considerable effort, in itself
decreasing the fidelity of the teleportation process and overcomplicating
the present protocol.

We thank the VPG/Vice-Reitoria de P\'{o}s-Gradua\c{c}\~{a}o e Pesquisa-UCG
(NGA), CAPES(SS, WBC, IPQ), and CNPq (NGA), Brazilian Agencies, for the
partial support.

\end{document}